\documentclass[]{elsart}

\usepackage{psfig,graphicx}
\usepackage[figuresright]{rotating}

\newcommand{\seff}{\sin^2\theta^{\mbox{\footnotesize lept}}_{\mbox{\footnotesize eff}}}
\newcommand{\Seff}{\sin^2\theta_{\mbox{\footnotesize eff}}}

\begin{document}

  \begin{flushright}
    DESY 06-057\\
    ZH-TH 11/06
  \end{flushright}

  \begin{frontmatter}
  
    \title {Bosonic Corrections \\ to The Effective Weak Mixing Angle at
    ${\mathcal O}(\alpha^2)$} 

    \author{M. Awramik$^{1,2}$} \author{M. Czakon$^{3,4}$} \author{A. Freitas$^5$}

    \address{$^1$II. Institut f\"ur Theoretische Physik,
      Universit\"at Hamburg, \\ Luruper Chaussee 149, D-22761 Hamburg,
      Germany}

    \address{$^2$ Institute of Nuclear Physics, Radzikowskiego 152,
      PL-31342 Cracow, Poland}

    \address{$^3$ Institut f\"ur Theoretische Physik und Astrophysik, Universit\"at
      W\"urzburg, Am Hubland, D-97074 W\"urzburg, Germany}

    \address{$^4$ Department of Field Theory and Particle Physics,
      Institute of Physics, University of Silesia, Uniwersytecka 4,
      PL-40007 Katowice, Poland}
  
    \address{$^5$ Institut f\"ur Theoretische Physik,
	Universit\"at Z\"urich, \\ Winterthurerstrasse 190, CH-8057
	Z\"urich, Switzerland}

    \begin{abstract}
      We present the complete bosonic contributions to the effective
      weak mixing angle, $\seff$, at the two-loop level in the
      electroweak interactions.  We find their size to be about three
      times smaller than inferred from simple estimates from lower
      orders. In particular, for a Higgs boson mass, $M_H$, of $100$
      GeV they amount to $4 \times 10^{-6}$, and drop down by about an order
      of magnitude for $M_H = 200$ GeV. We estimate the intrinsic
      error of the theory prediction of $\seff$ to be $4.7 \times 10^{-5}$.
    \end{abstract}

  \end{frontmatter}
  
%%%%%%%%%%%%%%%%%%%%%%%%%%%%%%%%%%%%%%%%%%%%%%%%%%%%%%%%%%%%%%%%%%%%%%%%%%%%%%%%
%%%%%%%%%%%%%%%%%%%%%%%%%%%%%%%%%%%%%%%%%%%%%%%%%%%%%%%%%%%%%%%%%%%%%%%%%%%%%%%%
%%%%%%%%%%%%%%%%%%%%%%%%%%%%%%%%%%%%%%%%%%%%%%%%%%%%%%%%%%%%%%%%%%%%%%%%%%%%%%%%

  \section{Introduction}

  While no clear experimental evidence for the Higgs boson has been found so far,
  even today the Standard Model is accurately tested and the Higgs boson mass
  $M_H$ strongly constrained through precision measurements of $Z$- and $W$-boson
  properties. One of the most important observables in this context is the
  effective leptonic weak mixing angle $\seff$.
  It can be defined through the vertex form factors for the vector and
  axial-vector interactions between the $Z$ boson and fermions $f$:
  \begin{equation}
    \seff = \frac{1}{4} \left(1 - \Re \left( \frac{g_V(M_Z^2)}{g_A(M_Z^2)}
    \right) \right).
  \end{equation}
  with the $g_{V,A}$ the effective couplings in the vertex $i\; \overline{l}
  \gamma^\mu (g_V-g_A \gamma_5)l \; Z_\mu$.

  The experimental value for $\seff$ is derived from various asymmetries
  measured at the $Z$ resonance pole. The precision of the current
  experimental value  $\seff = 0.23153 \pm 0.00016$ \cite{lepewwg} could
  be improved by about one order of magnitude by a future linear
  collider experiment \cite{gigaz}. Since the radiative corrections to
  $\seff$ depend sensitively on the value of $M_H$, the high
  experimental precision allows to put strong constraints on the Higgs
  boson mass when the Standard Model is assumed to be valid. Thus a lot
  of effort has been put into accurate theoretical calculations for
  $\seff$. While one-loop corrections and two- and three-loop QCD
  corrections have been known for several years
  \cite{1loop,QCD2L,QCD3L}, only recently the fermionic two-loop
  corrections, i.e. the two-loop contributions with at least one closed
  fermion loop, were computed in
  \cite{fermionic1,fermionicRest1,fermionicRest2} and confirmed in
  \cite{fermionic2}. In addition, leading three-loop effects of
  order ${\mathcal O}(\alpha^3)$ and ${\mathcal O}(\alpha^2 \alpha_s)$
  for large values of the top quark mass $m_t$ have been calculated
  \cite{faisst}, as well as the behavior of the full ${\mathcal
    O}(\alpha^3)$ corrections for large $M_H$ \cite{radja}. Finally, the
  precision of the QCD corrections to the universal part, the $\Delta
  \rho$ parameter, has been pushed to the four-loop level
  \cite{four1,four2}.

  However, for the remaining bosonic two-loop corrections, only a
  partial result for the $M_H$-dependent diagrams is available so far
  \cite{Hollik:2005ns}.  The goal of this work is to finalize the
  calculation of the ${\mathcal O}(\alpha^2)$ two-loop corrections by
  giving a complete result for the bosonic two-loop contributions.

  At tree-level, the effective weak mixing angle is identical to the
  on-shell weak mixing angle $\sin^2 \theta_{\rm W} =
  1-M_W^2/M_Z^2$. The effect of higher-order corrections to the $Zll$
  vertex can be summarized in the quantity $\Delta\kappa$,
  \begin{equation}
    \seff = \left(1 - \frac{M_W^2}{M_Z^2} \right)  \left(1 +
    \Delta\kappa \right),
    \label{kappa}
  \end{equation}
  where for the purpose of this work, it is understood that $M_W$ and
  $M_Z$ are defined in the on-shell scheme. The most precise result for
  $\seff$ is obtained when using the Fermi constant $G_\mu$ instead of
  $M_W$ as input. Then the calculation of $\seff$ as a function of
  $G_\mu$ involves also the computation of the radiative corrections to
  the relation between $G_\mu$ and $M_W$. This has been carried out with
  complete electroweak two-loop corrections in
  \cite{muon1,muon2,MWFitting}.  In this letter, the remaining
  bosonic two-loop corrections to the form factor $\Delta\kappa$ are
  presented.

%%%%%%%%%%%%%%%%%%%%%%%%%%%%%%%%%%%%%%%%%%%%%%%%%%%%%%%%%%%%%%%%%%%%%%%%%%%%%%%%
%%%%%%%%%%%%%%%%%%%%%%%%%%%%%%%%%%%%%%%%%%%%%%%%%%%%%%%%%%%%%%%%%%%%%%%%%%%%%%%%
%%%%%%%%%%%%%%%%%%%%%%%%%%%%%%%%%%%%%%%%%%%%%%%%%%%%%%%%%%%%%%%%%%%%%%%%%%%%%%%%
 
  \section{Outline of the calculation}  

  Any higher order calculation consists of two parts: the computation
  of the bare diagrams and the determination of the renormalization
  constants. The latter has been discussed at length in connection to
  two-loop electroweak precision observables in \cite{muon1,muon2}
  (see also \cite{Jegerlehner:2001fb}). We are, therefore, left with
  the calculation of the bare diagrams, which in our case, are massive
  two-loop three-point functions with two massless and one massive
  external leg.

  Just as in the case of the fermionic corrections \cite{fermionic1},
  there are three mass scales in the problem, with the difference that
  there is no dependence on the  top quark, but on the Higgs boson
  mass. This is, however, an important difference, because contrary to
  $m_t$, $M_H$ is not a fixed parameter and can assume a broad range
  of values.  From the many possible strategies that one might
  apply, we chose to expand in the various parameters in order to
  obtain a result expressed through single scale integrals, which are
  in fact just numbers to be determined in a final step.

  In a first step, we apply an expansion in the difference of the
  masses of the $W$ and $Z$ bosons, where the expansion parameter is
  just $s_W^2$. Since there are diagrams where there is a threshold
  when $M_W = M_Z$, the appearance of divergences at higher orders in
  the expansion is inevitable.  In this case, we apply the method of
  expansions by regions, see \cite{smirnov}.  The two regions that
  contribute to the result come from the  {\it ultrasoft momenta},
  $k_{1,2} \sim s^2_W M_Z$,  and {\it hard momenta}, $k_{1,2} \sim
  M_Z$.  The new integrals that appear from this procedure are
  presented in  Ref.~\cite{bosonic1}, whereas the reduction to the set
  of master integrals proceeds with Integration-By-Parts identities
  \cite{ibp} solved with the Laporta algorithm \cite{laporta} as
  implemented in the {\it IdSolver} library \cite{idsolver}.

  The Higgs boson is treated in two regimes. For low masses we expand
  in the mass difference between $M_H$ and $M_Z$, with the expansion
  parameter defined to be

  \begin{equation}  
      s_H^2=1-\frac{M_H^2}{M_Z^2},
  \end{equation}
  
  where this time, no thresholds are encountered. To guarantee a reasonable
  precision, we compute six terms in the combined expansion in $s_W^2$
  and $s_H^2$. For the second regime, which is the region where $M_H
  \gg M_Z$, we apply the large mass expansion, Ref.~\cite{smirnov}. 

  The resulting single scale master integrals are treated with various
  methods, usually with two or three different ones for test purposes.
  Most can be obtained with numerical integration, using dispersion
  relations (see Method II in \cite{fermionicRest2}).  For diagrams of
  simpler topologies we use differential equations
  \cite{Kotikov:1991hm,Remiddi:1997ny} and large mass expansions,
  whereas for more complicated ones we used the {\tt MB} package
  \cite{Czakon:2005rk} implementing Mellin-Barnes methods
  \cite{Smirnov:1999gc,Tausk:1999vh} (see also
  \cite{Anastasiou:2005cb}). Whenever possible we performed cross
  checks with sector decomposition, Ref.~\cite{sectors}.

  A final, algebraic check of all the procedures is the cancellation of
  the dependence on the gauge parameter, which we verified for the first
  orders of the expansion.

%%%%%%%%%%%%%%%%%%%%%%%%%%%%%%%%%%%%%%%%%%%%%%%%%%%%%%%%%%%%%%%%%%%%%%%%%%%%%%%%
%%%%%%%%%%%%%%%%%%%%%%%%%%%%%%%%%%%%%%%%%%%%%%%%%%%%%%%%%%%%%%%%%%%%%%%%%%%%%%%%
%%%%%%%%%%%%%%%%%%%%%%%%%%%%%%%%%%%%%%%%%%%%%%%%%%%%%%%%%%%%%%%%%%%%%%%%%%%%%%%%

  \section{Results}

\begin{figure}
    \begin{center}
      \includegraphics[width=8cm,height=11cm,angle=270]{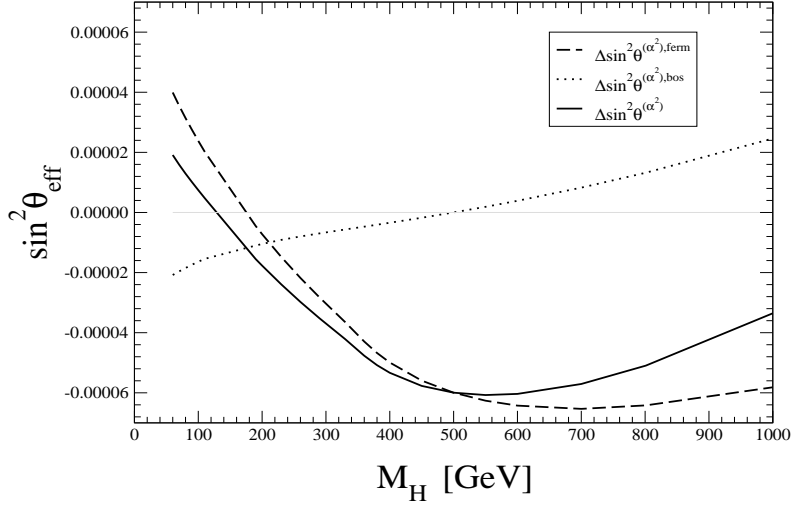}
      \caption{\label{cntbs2lEW}
	Two-loop electroweak contributions to the effective weak mixing angle, 
	with a fixed W boson mass as in Tab.~\ref{input}.}
    \end{center}
  \end{figure}

  The different electroweak contributions to $\Delta\kappa$ are shown
  in Tab.~\ref{tab:resultsMWfixed}, under the assumption that the $W$
  boson mass is fixed at its experimental value given in
  Tab.~\ref{input} together with the remaining input parameters. Note
  that $\Delta\kappa^{(\alpha^2), \mbox{ferm}}$ denotes the two-loop
  contribution of the fermionic diagrams known from
  \cite{fermionic1,fermionic2}, whereas $\Delta\kappa^{(\alpha^2),
  \mbox{bos}}$ is our new result, namely the two-loop contribution of
  the bosonic diagrams. The complete prediction, $\Delta\kappa$,
  contains additionally other known corrections. In particular, the
  following have been taken into account (see also \cite{fermionic1}):
  one-loop electroweak corrections, QCD corrections to the one-loop
  prediction at  the two- \cite{QCD2L} and three-loop level
  \cite{QCD3L}, ${\mathcal O}(\alpha^2 \alpha_sm_t^4)$  and ${\mathcal
  O}(\alpha^3 m_t^6)$ corrections to $\Delta \rho$ \cite{faisst},  as
  well as leading reducible effects at ${\mathcal O}(\alpha^2
  \alpha_s)$ and ${\mathcal O}(\alpha^3)$. The exact $M_H$ dependence
  of the two-loop contributions to $\seff$, obtained from
  $\Delta\kappa$ by rescaling with $1-M_W^2/M_Z^2$, can also
  be read off from Fig.~\ref{cntbs2lEW}, which makes it even more
  apparent that both fermionic and bosonic corrections are of the same
  order for low to moderate Higgs boson masses.

  \begin{table}[!htb]
    \caption {\small Higgs boson mass dependence of $\Delta \kappa$ evaluated 
      with a fixed W boson mass as in Tab.~\ref{input}. The
      normalization factor is $10^{-4}$.}
    \vspace{0.2cm}
    \begin{tabular}{| c || c | c | c || c |}
      \hline\hline
      $M_H\left[GeV\right]$ &  $\Delta\kappa^{(\alpha)}$ &  $\Delta\kappa^{(\alpha^2),\mbox{ferm}}$ & $\Delta\kappa^{(\alpha^2),\mbox{bos}}$ & $\Delta\kappa$ \\
      \hline
      100   &  \hspace*{0cm} 413.325 \hspace*{0cm} &  \hspace*{0cm} 1.07 \hspace*{0cm}  &  \hspace*{0cm}-0.74\hspace*{0cm} & \hspace*{0cm}372.93\hspace*{0cm}  \\
      200   &  \hspace*{0cm} 394.023 \hspace*{0cm} &  \hspace*{0cm}-0.32 \hspace*{0cm}  &  \hspace*{0cm}-0.47\hspace*{0cm} & \hspace*{0cm}353.20\hspace*{0cm}  \\
      600   &  \hspace*{0cm} 354.060 \hspace*{0cm} &  \hspace*{0cm}-2.89 \hspace*{0cm}  &  \hspace*{0cm} 0.17\hspace*{0cm} & \hspace*{0cm}313.13\hspace*{0cm}  \\
      1000  &  \hspace*{0cm} 333.159 \hspace*{0cm} &  \hspace*{0cm}-2.61 \hspace*{0cm}  &  \hspace*{0cm} 1.11\hspace*{0cm} & \hspace*{0cm}295.11\hspace*{0cm}  \\
      \hline\hline

    \end{tabular}
    \label{tab:resultsMWfixed}
  \end{table}

  \begin{table}
    \caption{\label{input} Input parameters, taken from \cite{lepewwg,Eidelman:2004wy}.
    }
    \begin{tabular}{|l|l|}
      \hline \hline 
      input parameter & value\\ 
      \hline 
      $M_W$ & $80.404 \pm 0.0030{\rm \; GeV}$ \\
      $M_Z$ & $91.1876 \pm 0.0021{\rm \; GeV}$ \\
      $\Gamma_Z$ & $2.4952 {\rm \; GeV}$ \\
      $m_t$ & $172.5 \pm 2.3 {\rm \; GeV}$ \\
      $m_b$ & $4.85 {\rm \; GeV}$ \\
      $\Delta\alpha(M_Z^2)$ & $0.05907 \pm 0.00036$ \\
      $\alpha_s(M_Z)$ & $0.119 \pm 0.002$ \\
      $G_\mu$ & $1.16637 \times 10^{-5} {\rm \; GeV^{-2}}$ \\
      \hline 
    \end{tabular}
  \end{table}

  In order to partially cancel large perturbative effects and lower the
  sensitivity to input parameters, the $W$ boson mass is customarily
  replaced by its value determined from $\mu$ decay, or equivalently
  from the Fermi constant, $G_\mu$. The transition is made possible by
  the knowledge of all relevant corrections to muon decay as discussed
  in \cite{MWFitting}.

  In view of the reparametrization, let us define the size of the
  complete bosonic corrections as the difference between the complete
  prediction of $\seff$ and the same prediction, where pure two-loop
  electroweak bosonic diagrams have been omitted both in $M_W$ and in
  $\Delta\kappa$. A rough estimate of the effect can be read off
  Tab.~\ref{tab:resultsMWfixed}. For example, for $M_H=100$ GeV and
  the current input parameters, where $\mbox{sin}^2\theta_W = 0.2225$,
  the   contribution to $\seff$ from $\Delta\kappa$ is $- 0.16 \times
  10^{-4}$, whereas using our previous results,
  \cite{muon1,muon2,MWFitting},  we find that the contribution from
  $M_W$ amounts to $0.2 \times 10^{-4}$. The large cancellation in the
  sum gives just $0.04\times 10^{-4}$. A similar cancellation occurs
  for other values of $M_H$ over a wide range, as illustrated in
  Tab.~\ref{tab:resultSineff}, which also gives the complete
  prediction. It is important to note that, contrary to the fermionic
  corrections, which strongly depend on the value of the top quark
  mass, our result for the bosonic corrections is stable within the
  input parameter uncertainties and can be simply added to our
  fitting formula from \cite{fermionic1}, although the effect is
  clearly negligible.

  \begin{table}[!htb]
    \caption {\small Higgs boson mass dependence of $M_W$ and $\seff$ with
      $G_\mu$ as input parameter. Quantities with the superscript
      [ferm] do not contain the two-loop electroweak bosonic
      corrections, whereas those with the superscript [ferm+bos] do
      contain them. $\Delta\Seff^{\mbox{}}$ is the shift induced by
      the bosonic corrections as described in the text. The
      normalization factor for $\seff$ is $10^{-4}$, and $M_H$ and
      $M_W$ are given in GeV.}
    \vspace{0.2cm}
      \begin{tabular}{| c || c | c || c | c || c |}
	\hline\hline
	$M_H$ & $M_W^{\mbox{[ferm]}}$ & $M_W^{\mbox{[ferm+bos]}}$ & $\Seff^{\mbox{[ferm]}}$ & $\Seff^{\mbox{[ferm+bos]}}$ & $\Delta\Seff^{\mbox{}}$ \\
	\hline
	100  & 80.3694  & 80.3684  & 0.231434 & 0.231438  & 0.04 \\
	200  & 80.3276  & 80.3270  & 0.231769 & 0.231769  & 0.00 \\
	600  & 80.2491  & 80.2490  & 0.232322 & 0.232327  & 0.05 \\
	1000 & 80.2134  & 80.2141  & 0.232563 & 0.232574  & 0.12 \\
	\hline\hline
      \end{tabular}
    \label{tab:resultSineff}
  \end{table}

  As a partial check of our calculation we have also compared the
  Higgs boson mass dependence of our result with the one published in
  \cite{Hollik:2005ns}. Upon using the same input parameters, we have
  found excellent agreement as shown in Tab.~\ref{tab:comparison}.
 
  \begin{table}[!htb]
    \caption {\small Comparison of the Higgs boson mass dependence of $\Delta \kappa$ 
      with \cite{Hollik:2005ns} (with $M_W = 80.4260$ GeV and $m_t =
      178$ GeV taken from that work). The subscript ``sub'' denotes
      subtraction at $M_H=100$ GeV. The normalization factor is $10^{-4}$.
    \vspace{0.2cm}}     
    \begin{tabular}{|c || c | c |}
      \hline\hline
      $M_H\left[GeV\right]$& $\Delta\kappa^{(\alpha^2),\mbox{bos}}_{\mbox{sub}}$ & $\Delta\kappa^{(\alpha^2),\mbox{bos}}_{\mbox{sub}}$ \cite{Hollik:2005ns}  \\ \hline
      100  & 0      & 0     \\
      200  & 0.266  & 0.265 \\
      600  & 0.914  & 0.914 \\
      1000 & 1.849  & 1.849 \\
      \hline\hline
    \end{tabular}
    \label{tab:comparison}
  \end{table}

%%%%%%%%%%%%%%%%%%%%%%%%%%%%%%%%%%%%%%%%%%%%%%%%%%%%%%%%%%%%%%%%%%%%%%%%%%%%%%%%
%%%%%%%%%%%%%%%%%%%%%%%%%%%%%%%%%%%%%%%%%%%%%%%%%%%%%%%%%%%%%%%%%%%%%%%%%%%%%%%%
%%%%%%%%%%%%%%%%%%%%%%%%%%%%%%%%%%%%%%%%%%%%%%%%%%%%%%%%%%%%%%%%%%%%%%%%%%%%%%%%

  \section{Discussion}

  Our calculation shows that the last piece of the two-loop
  electroweak corrections to the effective weak mixing angle, the one
  coming from purely bosonic diagrams, gives a very small
  contribution. In fact, being of the order of a few times $10^{-6}$, it is
  below the anticipated precision of the linear collider, not
  even to mention the current experimental accuracy. This strengthens
  the validity of our fitting formula \cite{fermionic1}. Furthermore,
  since the recent calculation of the ${\mathcal O}(G_\mu m_t^2
  \alpha_S^3)$ corrections to the rho parameter \cite{four1,four2},
  has also given a very small contribution, the results of
  \cite{fermionic1} implemented in {\tt ZFITTER} \cite{ZFITTER} still
  provide a reliable prediction for $\seff$ under consideration of all 
  new results.

  We estimate the error from unkown higher order corrections on
  $\seff$ as  described in \cite{fermionic1}, including contributions
  for the next missing loop orders, {\it i.e.} ${\mathcal
  O}(\alpha^3)$, ${\mathcal O}(\alpha^2 \alpha_s)$, ${\mathcal
  O}(\alpha \alpha_s^3)$.\footnote{In Ref. [6] we accidentally
  mentioned a number  for  the ${\mathcal O}(\alpha^2 \alpha_s^2)$
  contributions, but did not include it in  the combined error.} We
  find a total theoretical error of $4.7 \times 10^{-5}$,  which should
  be taken as a very conservative error estimate, which makes it
  however necessary to determine further missing corrections if we
  want to obtain a result at the level needed by a future linear
  collider.

%%%%%%%%%%%%%%%%%%%%%%%%%%%%%%%%%%%%%%%%%%%%%%%%%%%%%%%%%%%%%%%%%%%%%%%%%%%%%%%%
%%%%%%%%%%%%%%%%%%%%%%%%%%%%%%%%%%%%%%%%%%%%%%%%%%%%%%%%%%%%%%%%%%%%%%%%%%%%%%%%
%%%%%%%%%%%%%%%%%%%%%%%%%%%%%%%%%%%%%%%%%%%%%%%%%%%%%%%%%%%%%%%%%%%%%%%%%%%%%%%%

  \begin{ack}
    The work of M.A. was supported by the BMBF grant No 04-160.
    The work of M.C. was supported by the Sofja
    Kovalevskaja Award of the Alexander von Humboldt Foundation
    sponsored by the German Federal Ministry of Education and
    Research.

  \end{ack}

%%%%%%%%%%%%%%%%%%%%%%%%%%%%%%%%%%%%%%%%%%%%%%%%%%%%%%%%%%%%%%%%%%%%%%%%%%%%%%%%
%%%%%%%%%%%%%%%%%%%%%%%%%%%%%%%%%%%%%%%%%%%%%%%%%%%%%%%%%%%%%%%%%%%%%%%%%%%%%%%%
%%%%%%%%%%%%%%%%%%%%%%%%%%%%%%%%%%%%%%%%%%%%%%%%%%%%%%%%%%%%%%%%%%%%%%%%%%%%%%%%

\end{document}